\documentclass[3p,12pt,sort&compress]{elsarticle}

\usepackage{graphicx,amsmath,amssymb,amsfonts,latexsym,color,dcolumn,bm,subfigure,ulem}

\makeatletter
\def\ps@pprintTitle{%
 \let\@oddhead\@empty
 \let\@evenhead\@empty
 \def\@oddfoot{}%
 \let\@evenfoot\@oddfoot}
\makeatother

\usepackage{latexsym}
\usepackage{hyperref}
\usepackage{cleveref}
\usepackage{placeins}
\newcolumntype{L}{>{\centering\arraybackslash}m{6.5cm}}
\newcolumntype{K}{>{\centering\arraybackslash}m{13cm}}

\begin{document}

\renewcommand*{\thefootnote}{\fnsymbol{footnote}}

\title{Indications of an unexpected signal associated with the GW170817  binary neutron star inspiral\footnote{{\bf Journal Reference}: Astroparticle Physics 103 (2018) 1Ð6}}

\author[Purdue]{E. Fischbach}
\ead{ephraim@purdue.edu}
\author[Purdue]{V. E. Barnes}
\author[Berkeley,Purdue]{N. Cinko}
\author[Purdue]{J. Heim}
\author[Maryland,Purdue]{H. B. Kaplan}
\author[Wabash,Purdue]{D. E. Krause}
\author[Purdue]{~J.~R.~Leeman}
\author[NRL]{S. A. Mathews}
\author[Purdue]{M. J. Mueterthies}
\author[UCLA,Purdue]{D. Neff}
\author[Purdue]{M. Pattermann}
\address[Purdue]{Department of Physics  and Astronomy, Purdue University, West Lafayette, IN 47907, USA}
\address[Berkeley]{Department of Physics, University of California, Berkeley, CA 94720, USA}
\address[Maryland]{Department of Physics, University of Maryland, College Park, MD, 20742, USA}
\address[Wabash]{Physics Department, Wabash College, Crawfordsville, IN 47933, USA}
\address[NRL]{US Naval Research Laboratory, Material Science Division, Washington, DC 20375, USA}
\address[UCLA]{Department of Physics and Astronomy, University of California, Los Angeles, CA 90024, USA}

\begin{abstract}
We report experimental evidence at the 2.5$\sigma$ level for an unexpected signal associated with the GW170817 binary neutron star inspiral. This evidence derives from a laboratory experiment simultaneously measuring the $\beta$-decay rates of Si-32 and Cl-36 in a common detector. Whereas the Si-32 and Cl-36 decay rates show no statistical correlation before or after the inspiral, they are highly correlated ($\sim 95\%$) in the 5 hour time interval immediately following the inspiral. If we interpret this correlation as arising from the influence of particles emitted during the inspiral, then we can estimate the mass $m_{x}$ of these particles from the time delay between the gravity-wave signal and a peak in the $\beta$-decay data. We find for particles of energy 10~MeV, $m_{x}$ $\lesssim$ 16~eV which includes the neutrino mass region $m_{\nu}$ $\lesssim$ 2~eV.  The latter is based on existing limits for the masses $m_{i}$ of the three known neutrino flavors. Additionally, we find that the correlation is even stronger if we include data in the 80 minute period before the arrival of the gravity wave signal. Given the large number of radionuclides whose decays are being monitored at any given time, we conjecture that other groups may also be in a position to search for statistically suggestive fluctuations of radionuclide decay rates associated with the GW170817 inspiral, and possibly with other future inspirals.

\end{abstract}


\maketitle

\section{Introduction}

The recent  detection by the Advanced LIGO and Advanced Virgo detectors of a gravitational wave signal from a binary neutron (NS-NS) star inspiral \cite{Abbott 1,Abbott 2}, GW170817 was quickly followed by an optical detection of the remnants of the inspiral as well.  The gravitational wave signal was received at 12:41:04 UTC on 17 August 2017, and the source of the inspiral was localized to a luminosity distance of $1.30 \times 10^{8}$~ly = $1.23 \times 10^{26}$~cm from Earth.  A subsequent search for high-energy neutrinos associated with the inspiral by the $\textsc{antares}$, IceCube, and Pierre Auger observatories did not detect any neutrinos \cite{Albert}.

Here we report a possible indication of an additional signal associated with the NS-NS inspiral obtained from an experiment at Purdue University that was underway at the time of the inspiral. The experiment was aimed at obtaining an improved half-life for Si-32 ($T_{1/2} \sim 172$~yr), using an upgraded apparatus originally designed and built by a group at Brookhaven National Laboratory \cite{Alburger,Harbottle}, and subsequently donated to our group. The special feature of this apparatus, shown  in Fig. 1, is that it allowed the decay rate of Si-32 to be continually compared to that of Cl-36 ($T_{1/2} \sim 301,000$~yr) which acted as an internal calibration standard. As noted in \cite{Alburger}, Si-32 decays via $\beta^{-}$ emission 100\% of the time to the ground state (g.s.) of P-32 emitting an electron with energy $Q(\beta^{-})= 225$ keV. P-32 ($T_{1/2} = 14.28$~d) decays in turn 100\% of the time via $\beta^{-}$ emission to the g.s. of S-32 with $Q(\beta^{-}) = 1709$~keV. Cl-36 ($T_{1/2} \sim 301,000$~yr) decays via $\beta^{-}$ emission to the g.s. of Ar-36 98.1\% of the time with $Q(\beta^{-}) = 709$~keV, and via electron capture (e.c.) to the g.s. of S-36 1.9\% with $Q(\mbox{e.c.}) = 1144$~keV. Since the 225 keV $Q(\beta^{-}$) for Si-32 decay is too low an energy to be detected reliably, the Si-32 decay rate is determined from the subsequent decay of P-32 which is presumed to be in equilibrium with Si-32. Any short-term perturbations of the ``Si-32'' decay rate would actually be perturbations of the P-32 decay rate.

The Si-32 and Cl-36 samples were separately positioned on top of two pedestals affixed to two precision micrometers, one of which (B) is shown in Fig. 1. 
\begin{figure}[t]
\centering\includegraphics[width=4in]{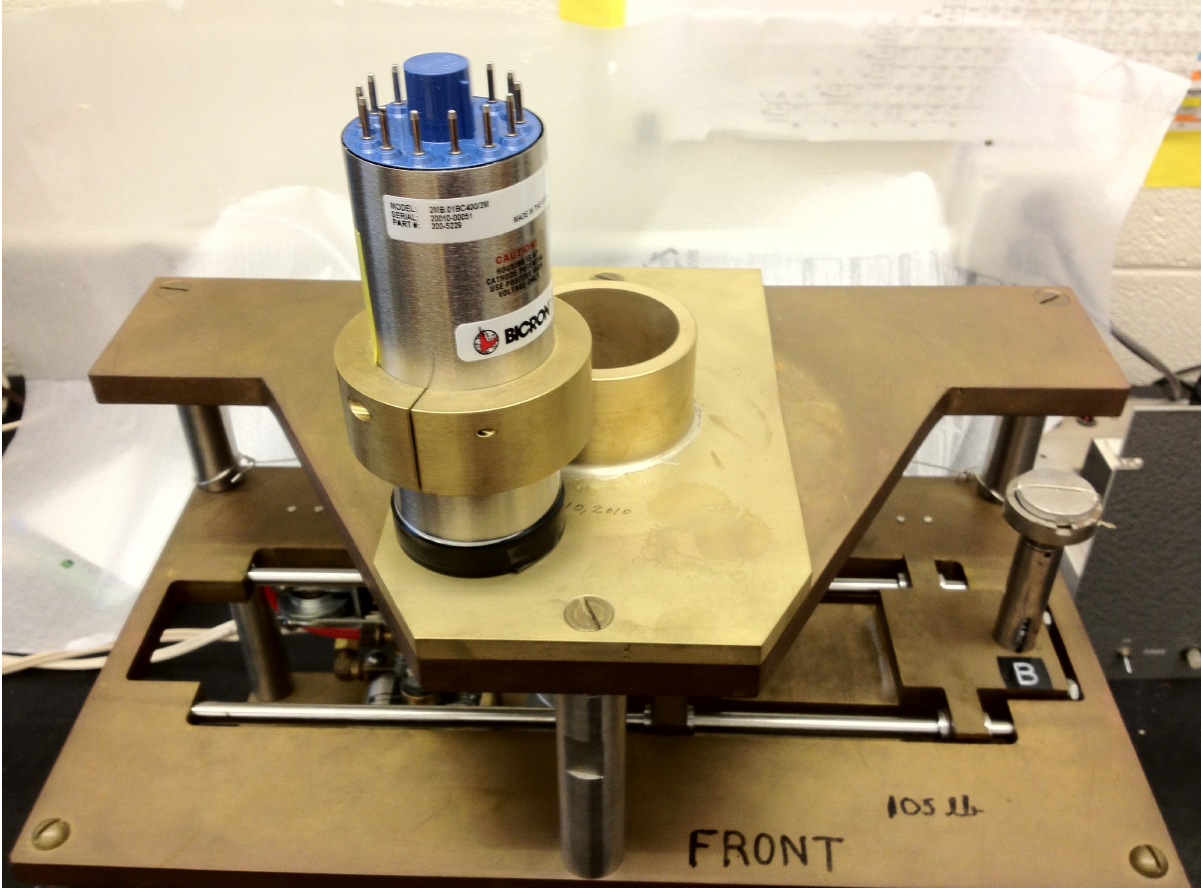}
\caption{The BNL apparatus for measuring the decay rates of Si-32 and Cl-36. Shown is pedestal B on which the Si-32 sample is placed, the Bicron plastic scintillation detector, and the collar to keep the detector rigidly in place. In the configuration shown, the Cl-36 sample (hidden) would be directly under the detector. The samples are driven along the track rails in the foreground by a pneumatic system which places each in turn under the detector. See text for further details.}
\label{apparatus figure}
\end{figure}
The micrometers were rigidly fixed to a carriage which moved back and forth  to position each sample under the detector(shown) in 30 minute intervals. Given the long half-life of Cl-36, its measured rate should have been effectively constant over the 4-year duration of the original experiment ($4/301,000 \simeq 1.3 \times 10^{-5}$) \cite{Alburger}. Hence any short-term or long-term deviations from a constant Cl-36 count rate could be attributed to drifts in the apparatus, which would then apply to the Si-32 count rate as well. By comparing the Si-32 and Cl-36 count rates over time, specifically by computing their ratio Si-32/Cl-36, the correct Si-32 half-life could be determined.


The BNL experiment was the first to report an unexpected annual oscillation in the decay rates of their samples, superimposed on their respective exponential decay curves \cite{Alburger}.  Data acquired during approximately the same period at the Physikalische Technische Bundesanstalt (PTB) in Germany exhibited a similar annual variation in their decay data for Ra-226 \cite{Siegert}.  Subsequently a number of other decay experiments and analyses have reported similar results \cite{Flkenberg,Fischbach 2,Javorsek,Jenkins 1,Jenkins 2,Jenkins 3,Norman,OKeefe,Parkhomov,Sturrock 1,Sturrock 2,Sturrock 3,Sturrock 4}.  Of particular significance in the present context is evidence of changes in nuclear decay rates associated with solar storms \cite{Jenkins 1,Mohsinally}.  Although there is at present no known mechanism to account for these correlations, various considerations suggest that they may be attributed to solar or relic neutrinos, or to similar dark matter candidates.  If we assume this to be the case, then it is natural to ask whether a signal from the GW170817 inspiral might also show up in our system.

\section{Experimental Details}

As we now demonstrate, we have applied a variant of the BNL technique to compare the short-term behaviors of the Si-32 and Cl-36 count rates before, during, and after the inspiral event.  We then show that in the 5-hour period immediately following the arrival of the inspiral signal, the Si-32 and Cl-36 signals become highly correlated, which we interpret as evidence for an external influence affecting the respective decays.  Our modified BNL detection system is shown in Fig. 1.  We upgraded the original BNL apparatus by introducing a collar (shown) designed to secure our detector to the sample changer.  The data acquisition system (DAQ) consisted of a 2-inch Bicron plastic scintillation detector (shown) coupled to an Ortec~276 base and preamplifier.  The remainder of the system consisted of a spectroscopy amplifier, and a multi-channel analyzer (MCA) connected to a PC running Maestro 32 and custom LabView software.  The upgraded sample changer was placed in a  modified 55 gallon steel drum  which acted as Faraday shield, and was kept at a controlled pressure with dry nitrogen.  This arrangement also allowed us to minimize variations in relative humidity.  The temperature in the room was controlled by a water-cooled air conditioner set to $21^{\circ}$C.  The entire system was then completely encased in a styrofoam box with 2-inch thick walls to suppress temperature variations.  The temperature inside the box itself was further controlled by a Technology, Inc. model AC-073  heating/cooling unit set to $22.0^{\circ}$C.  Electrical power to the system was supplied by a constant voltage transformer to mitigate any effects from fluctuating line voltage.  This system allowed the voltage output to be held constant at 117~V for an input voltage in the range 109--125~V.  Further details can be found in \cite{Heim}. Since the Si-32 and Cl-36 counts are acquired by the BNL system in successive 30 minute runs, the data acquired from one sample should be statistically uncorrelated to the data acquired from the other. Hence, in the absence of evidence for a malfunction in the detection system, any indication of a correlation can be attributed to either a statistical fluctuation (see below), or to evidence of an external influence common to both the Si-32 and Cl-36 samples.
\begin{figure}[t]
\centering\includegraphics[width=4in]{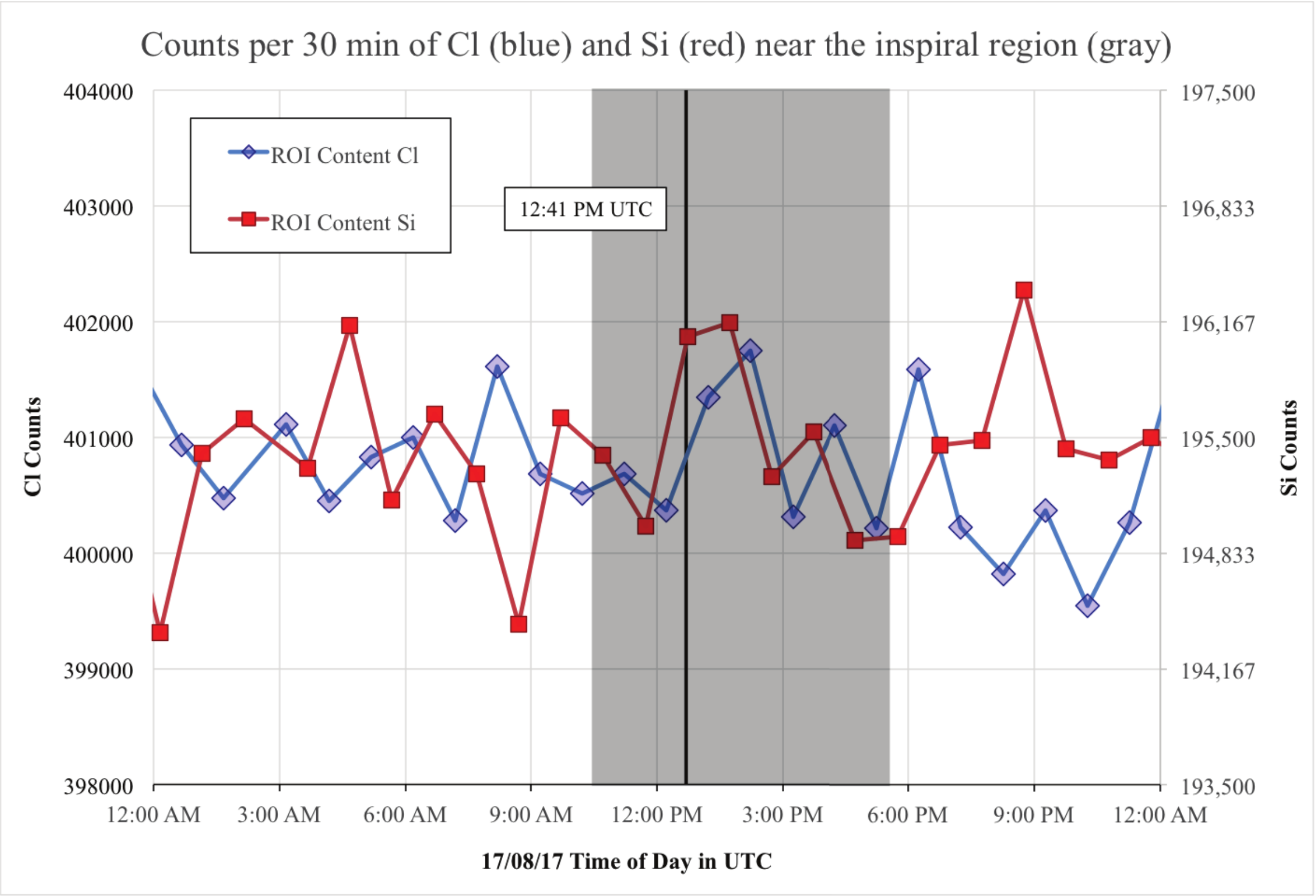}
\caption{Si-32 and Cl-36 count rates during August 2017. The scales on the respective vertical axes are arbitrary, and do not affect the determination of correlations between the decay rates. The shaded gray region encompasses 7 pairs of Si-32/Cl-36 counts which our analysis suggests are highly correlated, and the dark vertical line at 12:41 UTC on 17 August 2017 denotes the GW170817 gravity wave's time of arrival.}
\label{data graph figure}
\end{figure}
\begin{figure}[t]
\centering\includegraphics[width=4in]{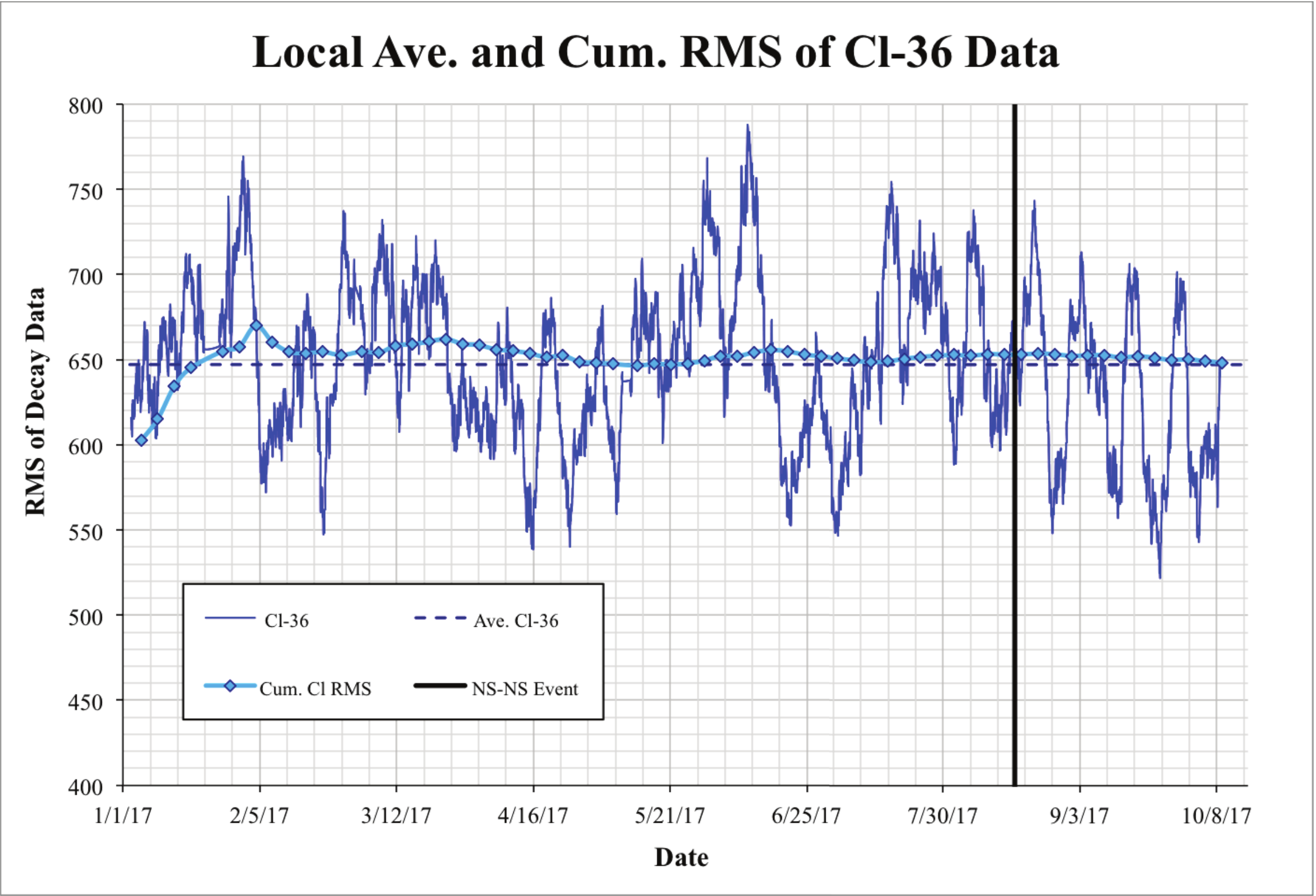}
\caption{The local average and cumulative root-mean-square (RMS) fluctuations of the 2017 Cl-36 decay data for the period 1 January through 11 October. Each of the 6,438 points represented in the figure is the 101 point rolling RMS average deviation from the mean, including that point and the 50 earlier and later points. The horizontal dashed line denotes the average number of counts for the entire data set, where $\overline{N} = 400,864$ and $\sqrt{\overline{N}} = 632$. Also shown are both the individual RMS values, whose average is 649, and the cumulative RMS values as a function of time. See text for further details.}
\label{apparatus figure}
\end{figure}
In what follows we verify that the fluctuations in the decays of the Si-32 and Cl-36 are in fact statistically uncorrelated, as expected, with three significant exceptions, one of which happens during the period immediately preceding and following the inspiral. To facilitate the various statistical tests that we have carried out, the Si-32 data have been correlated with the Cl-36 data points which are 30 minutes later in time. As we have noted above, we expect that the Si-32 and Cl-36 decay fluctuations should be uncorrelated, precisely because they are not acquired by our detection system at the same time. Matching the Si-32 data to the Cl-36 data points 30 minutes earlier in time does not produce a high correlation (see Table 1).  This asymmetry might be due to the start of the main signal having arrived when the Si-32 source was being counted and the Cl-36 was not under observation. 

\section{Experimental Results}

In Fig. 2 the dark vertical line denotes the time 12:41 UTC on 17 August 2017 when the LIGO/Virgo detectors observed a gravity-wave signal from a NS-NS inspiral.  We define the ``post-inspiral region'' to be the subsequent 5 hour portion of the shaded in gray region containing the 5 Si-32 and 5 Cl-36 sets of 30 minute counts.  To test for the statistical significance of the peak centered in this region we have carried out a Pearson correlation test on the 5 sets of Cl-36 and Si-32 points in the post-inspiral region.  We find the Pearson correlation coefficient to be 0.950, corresponding to a (one-sided) false-alarm probability of $6.7 \times 10^{-3}$ for positive correlations, as shown in Table 1. In contrast, the same test applied to the similarly shifted data set for the entire month of August 2017 (excluding the post-inspiral region) gives a Pearson correlation coefficient of $-4.2 \times 10^{-3}$.  This result is, of course, what would be expected for two uncorrelated data sets experiencing independent random $\sqrt{N}$ fluctuations.

To confirm the preceding estimate of the statistical significance of the correlation highlighted in Fig. 2, we have carried out an analogous search for other highly positively-correlated  fluctuations in calendar 2017 from 1 January to 11 October having a statistical significance $\geq 0.949$. As noted above, each day typically produces 24 Si-32 and 24 Cl-36 runs each of 30 minutes in duration. Hence the 284 day period should produce 6,816 Si-32/Cl-36 pairs of data to which we can apply an analysis identical to what we carried out for data acquired during the August 17 post-inspiral region. After excluding periods during which our apparatus was offline, we found 6,440 other regions to which we could apply the same analysis used in the post-inspiral region. Of these, 34 had correlation coefficients $\ge$ +0.949, corresponding to false-alarm probabilities of $\leq 6.7 \times 10^{-3}$. The agreement between this value, and the estimate for the 17 August 2017 5-hour post-inspiral region supports our estimate of the statistical significance of the post-inspiral correlation.

As a further check on the stability of our detection system we verify that the root-mean-square (RMS) fluctuations of the Cl-36 data outside the inspiral region are in accord with the usual ``$\sqrt{N}$ rule'' for statistically uncorrelated data.  The average number of Cl-36 counts $\overline{N}$ in a 30 minute interval for the period 1 January through 11 October is $\overline{N} = 400,864$ (again excluding the gray region in Fig. 2). We first calculate the rolling RMS deviation 
\begin{equation}
\sigma_i = \sqrt{ \hspace{0.5ex} \sum_{j=i-50}^{i+50} \hspace{0.8ex} \frac{(N_j-\overline{N})^{2}}{101} \hspace{1.2ex}} \nonumber
\end{equation}
and then the average $\langle \sigma_i \rangle$ over the data set (see Fig. 3). As seen from the figure, $\langle \sigma_i \rangle = 649$, to be compared with the expected value $\sqrt{\overline{N}} = 632$ assuming only statistical fluctuations. An RMS
test to the Si-32 data requires comparing counts to a fitted curve rather than to a simple average, due to the shorter half life.  For the Si-32 data we find an RMS of 455 counts, quite close to the square root of the number of counts ($\overline{N} = 199,000$), which is 446. The close agreement between $\langle \sigma_i \rangle$ and $\sqrt{\overline{N}}$ suggests that the fluctuations in both the Cl-36 and Si-32 decay rates outside the inspiral region are in fact statistical and not systematic.

\section{Experimental Implications}

If we interpret the features in the Si-32/Cl-36 data as arising from a flux of particles of mass $m_{x}$ associated with the NS-NS inspiral, then we can estimate $m_{x}$ from the difference in the arrival times of these features and the gravity wave signal. Let $(t_{g} + \delta t)$ denote the time-of flight of the particles of mass $m_{x}$, where $t_{g} = 1.30 \times 10^{8}$~yr is the time-of-flight of the gravitational wave presumed to be traveling at the speed of light \cite{Abbott 3}. From Fig. 2 we estimate that $\delta t$ ranges from 1.5 hours down to possibly very few minutes. Note that the data points are plotted at the completion times of each half-hour of data. Note also that one must make assumptions about the duration and timing of the emission of particles relative to the final chirp as the neutron stars coalesce. Using relativistic kinematics we find $E_{x}=m_{x}(1-v_{x}^{2})^{-1/2}$ (with $c = 1$) and hence
\begin{equation}
(m_{x}/E_{x})^{2} = 1 - v_{x}^2 \simeq 2 \delta t/t_{g}.
\label{m E}
\end{equation}
Combining Eq.~(\ref{m E}) with $t_{g}$ and the preceding value of $\delta t$ = 1.5 hours we find
\begin{equation}
m_{x}/E_{x} < 16\times10^{-7}.
\label{m/E}
\end{equation}
If we were to further assume that $E_{x}$,  the energy of the emitted neutrinos, or neutrino-like particles ``neutrellos'' \cite{Fischbach 3}, is $E_{x}\simeq 10$~MeV, we then find from Eq.~(\ref{m/E}), 
\begin{equation}
m_{x} < 16~{\rm eV}.
\label{mx}
\end{equation}

We note from Eq.~(\ref{m E}) that the $\sim 5$ hour spread in the arrival times of whatever particles might be responsible for the observed features may be attributable in part to a spread in the energies $E_{x}$ in Eq.~(\ref{m E}). This energy spread could arise in part from the scattering of the emitted particles leaving the binary. Furthermore, if these particles are in fact neutrinos, or neutrino-like "neutrellos" \cite{Fischbach 3}, then the expected flavor oscillations could also somewhat broaden the spread of arrival times compared to that of the gravity signal.

The result for $m_{x}$ in Eq.~(\ref{mx}) is larger than the limits inferred in several scenarios for the presumed neutrino mass eigenstates $m_{1}$, $m_{2}$, and $m_{3}$. These are derived from neutrino oscillation data combined with limits on the mass of $\bar{\nu}$\textsubscript{e} from tritium $\beta$-decay, $m(\bar{\nu}$\textsubscript{e}) $\lesssim$ 2 eV  \cite{Olive}:
\begin{equation}
\mbox{Normal Hierarchy:}~m_{1} \ll m_{2} < m_{3}: m_{2} \simeq 0.0087~{\rm eV}; m_{3} \simeq 0.05~{\rm eV},
\label{normal}
\end{equation}
\begin{equation}
\mbox{Inverted Hierarchy:}~m_{3} \ll m_{1} < m_{2}: m_{1,2} \simeq 0.049~{\rm eV},
\label{inverted}
\end{equation}
\begin{equation}
\mbox{Quasi-degenerate:}~
m_{1} \simeq m_{2} \simeq m_{3} \gtrsim 0.1~{\rm eV}.
\label{degenerate}
\end{equation}
Even the largest result Eq.~(\ref{degenerate}), which is compatible with the limit $m_{\nu} \geq  0.4$~eV derived from the stability of neutron stars and white dwarfs \cite{Fischbach 1}, would lead to a neutrino mass limit smaller than that obtained for $m_{x}$ in Eq.~(\ref{mx}). However, if we were to assume that $E_{x}\simeq 1$~MeV in Eq.~(\ref{m/E}), which is the characteristic neutrino energy in neutron $\beta$-decay, we would find $m_{x} \simeq (1.6 \pm 0.3)$~eV, which could be compatible with existing neutrino data in models involving sterile neutrinos.  Additionally, taking into account the earliest part of the signal further reduces the mass estimate, as could inclusion of the quoted uncertainties in $t_g$ \cite{Abbott 1}. For other neutrino mass limits refer to \cite{Giusarma,Vagnozzi,Bilenky}.

The high end of the mass estimates does not favor known neutrinos as the agents.  A further issue is a comparison of the estimated fluence of neutrinos reaching the Earth from the NS-NS inspiral versus the known fluence of solar neutrinos reaching the Earth during the 5-hour event region shown in Fig. 2. We note from \cite{Abbott 2} that the characteristics of neutrinos emitted from the NS-NS inspiral are expected to be similar to those arising from core collapse in a supernova for which estimates are available. We assume, as representative numbers, a neutrino luminosity $L = (1$--$3) \times 10^{53}$ erg/s lasting for $\sim$10~s. If the typical emitted neutrino energy is $\sim 10$~MeV, then the number $N$ of emitted neutrinos is $N \simeq (6$--$18) \times 10^{58}$. The resulting fluence $F_{\rm NS}$ at the Earth at a distance $d = 1.23 \times 10^{26}$ cm is then $F_{\rm NS} \simeq (3$--$10) \times 10^{5}$ $\nu$/cm$^{2}$. By way of comparison, the total fluence $F_{\odot}$ of neutrinos from the Sun during the 5-hour inspiral region shown in Fig. 2 would be approximately  
$F=(65 \times 10^{9}~\nu/{\rm cm}^{2}\cdot {\rm s})(1.8 \times 10^{4}~{\rm s})\simeq 1 \times 10^{15}~\nu/{\rm cm}^{2}$, and hence $F_{\rm NS} \ll F_{\odot}$.

The preceding analysis has focused on the 5 hour period following the arrival time of the gravity wave, on the assumption that no signal from the inspiral could exceed the speed of the gravity wave and hence precede it. However, a close examination of Fig. 2 reveals that the two additional pairs of points taken just before the gravity wave arrived, are also highly correlated. We have carried out an analysis of the full 7 hour region including the early data points, defined by the full gray shaded region in Fig.~2, following the same steps as described previously for the 5-hour post-inspiral region. We find for the 7-hour shaded region a Pearson correlation coefficient of 0.954, compared to 0.950 for the 5-hour post inspiral region, as listed in Table~1. This corresponds to a false-alarm probability of $4.3 \times 10^{-4}$. By way of comparison, the 7-hour Pearson correlation coefficient for the entire month of August 2017 implies a false alarm probability of  $-4.2 \times 10^{-3}$. We have also determined that for the period 1 January--11 October there are a total of 6,438 possible 7-hour pairs (again excluding times when our system was offline). Of these only 3 gave Pearson coefficients $\geq +0.954$, corresponding to false alarm probabilities of $\leq 4.3 \times 10^{-4}$. In addition to the inspiral event on 17 August, there were high correlation periods for the shifted 7-hour data on 25 May and on 15 June, which could be attributed to expected random fluctuations of the data.

\section{Conclusions}

To summarize, the design of the BNL detection system ensures that the Si-32 and Cl-36 decays should be completely uncorrelated absent any external driving signal. However, what we have shown is that our system detected a correlation on 17 August 2017 of decay rates, with an upward fluctuation which peaked at 93 minutes following the arrival of the gravity wave detected by LIGO. There were no observed correlations between this feature and any fluctuations in temperature, pressure, humidity or line voltage recorded in our system. The five-hour correlation false-alarm probability of the inspiral event was $6.5 \times 10^{-3}$, as determined from statistical tests we carried out for the period 1 January--11 October 2017. Combining our data with the distance to the inspiral binary as determined by the LIGO/Virgo analysis allows us to infer that our signal may be attributable to the arrival of particles of mass possibly considerably less than $\sim(10$--$20)$~eV, which could be compatible with neutrino mass limits determined from current laboratory experiments and astrophysical arguments. Our results may also be compatible with a scenario in which part of the energy of the inspiral is emitted in the form of some new particles. However, the ``Si/Cl event'' confronts us with other puzzles: In the first place, the fluence of neutrinos, or hypothetical neutrino-like ``neutrellos'' \cite{Fischbach 3}, appears to be too small to account for the observed data, at least in the framework of existing models of supernova formation or binary neutron star inspirals. Additionally, the extra correlation seen by including the pre-merger data points, while increasing the apparent significance of the 7-hour correlation to the 3.3$\sigma$ level, may of course be just a statistical fluctuation.  There is no presently known process where large fluxes of energy might emerge many minutes before the final NS merger. However, we note that data acquired at the Mont Blanc Underground Neutrino Observatory (UNO) may have also detected a precursor signal associated with the SN1987A event \cite{Aglietla,Ehrlich}. Finally, the improbability of even the 5-hour correlation's coinciding with the arrival of the gravity wave leads us to note that there may exist many other experiments studying radioactive decays which could have detected a similar signal, and we encourage other groups to examine their data.  This may make it possible to acquire additional supporting data from future NS-NS inspirals. 
\begin{table}[t]
\begin{center}
\begin{tabular}{ | L | L | }
\hline
\multicolumn{2}{ | K | }{Intervals of 5 Cl-Si Pair Measurements (right-shifted)} \\
\hline
\hline
Region of Interest & Pearson Coefficient $r$ \\
\hline
NS-NS Inspiral & 0.94993 \\
\hline
2017 Calendar to 11 Oct & 0.02526 \\
\hline
2017 w/o Inspiral & 0.02516 \\
\hline
\end{tabular}

\begin{tabular}{ | L | L | }
\hline
\multicolumn{2}{ | K | }{Intervals of 5 Cl-Si Pair Measurements (left-shifted)} \\
\hline
\hline
Region of Interest & Pearson Coefficient $r$ \\
\hline
NS-NS Inspiral & 0.28341 \\
\hline
2017 Calendar to 11 Oct & 0.03733 \\
\hline
2017 w/o Inspiral & 0.03729 \\
\hline
\end{tabular}

\begin{tabular}{ | L | L | }
\hline
\multicolumn{2}{ | K | }{Intervals of 7 Cl-Si Pair Measurements (right-shifted)} \\
\hline
\hline
Region of Interest & Pearson Coefficient $r$ \\
\hline
NS-NS Inspiral & 0.95383 \\
\hline
2017 Calendar to 11 Oct & 0.02526 \\
\hline
2017 w/o Inspiral & 0.02502 \\
\hline
\end{tabular}

\begin{tabular}{ | L | L | }
\hline
\multicolumn{2}{ | K | }{Intervals of 7 Cl-Si Pair Measurements (left-shifted)} \\
\hline
\hline
Region of Interest & Pearson Coefficient $r$ \\
\hline
NS-NS Inspiral & 0.05919 \\
\hline
2017 Calendar to 11 Oct & 0.03733 \\
\hline
2017 w/o Inspiral & 0.03729 \\
\hline
\end{tabular}
\end{center}
\caption{Pearson correlation coefficients $r$ relating the Si-32 and Cl-36 data sets for time intervals during the period 1 January -- 11 October 2017. The 5-hour region is the period to the right of the vertical black line in the shaded area of Figure 2 denoting the arrival of the NS-NS inspiral gravity wave. The designation (right-shifted) refers to the Si-32 data which have been shifted to the right (later) by 30 minutes, so as to align with the Cl-36 data. The 7-hour region encompasses the entire shaded area in Fig. 2. See text for further details.}
\end{table}

\section*{Acknowledgements}

The authors are deeply indebted to David Alburger and Garman Harbottle for generously donating their unique apparatus to the Purdue group, along with their Si-32 and Cl-36 samples.  We are also indebted to our many colleagues for helpful communications and insights, including Maxim Barkov, George Bashinzhagyan, Rob de Meijer, Don Eckhardt, Dimitrios Giannios, Mark Haugan, Andrew Longman, Maxim Lyutikov, Jonathan Nistor, Alexander Parkhomov, Jeff Scargle, Gideon Steinitz, and Peter Sturrock.

\end{document}